\documentclass[showpacs,preprintnumbers,amsmath,amssymb]{revtex4}
\usepackage{amsmath,amssymb,graphics,epsfig,subfigure}
\usepackage{color}

\begin{document}
\renewcommand{\baselinestretch}{1.3}

\title{Reduced thermodynamics and an exact phase transition of five-dimensional Kerr-AdS black holes with equal spin parameters}

\author{Shao-Wen Wei$^{1,2}$  \footnote{weishw@lzu.edu.cn},
        Yu-Xiao Liu$^{1}$  \footnote{liuyx@lzu.edu.cn},}

\affiliation{ $^{1}$Institute of Theoretical Physics $\&$ Research Center of Gravitation, Lanzhou University, Lanzhou 730000, People's Republic of China,\\
$^{2}$Department of Physics and Astronomy, University of Waterloo, Waterloo, Ontario, Canada, N2L 3G1}

\begin{abstract}
Multi-spinning higher dimensional Kerr-AdS black holes admit the stable small-large black hole phase transition of van der Waals type. In this paper, we study the exact critical phenomena and phase structure in five-dimensional spacetime. First, we examine the thermodynamic laws in the reduced parameter space and find that they are quite different from the conventional thermodynamic laws. Then based on the reduced laws, the phase structure in different parameter spaces is investigated. The stable and metastable black hole phases are clearly displayed. We present a highly accurate fitting formula for the coexistence curve of small and large black hole phases. Using this fitting formula, we examine the critical exponents when the black hole system approaches the critical point along the coexistence curve. Moreover, employing the dimensional analysis and symmetry analysis, we also give a numerical study of the critical point for the unequal spinning black holes. These results are very useful on further understanding the microstructure of the multi-spinning black holes in higher dimensional spacetime.
\end{abstract}

\keywords{Black holes, critical phenomena, phase structure}

\pacs{04.70.Dy, 04.50.Gh, 05.70.Ce}

\maketitle

\section{Introduction}

Black hole thermodynamics in anti-de Sitter (AdS) space has been a great interest in black hole physics. Especially, the black hole phase transition continues to be a fascinating issue. Several decades ago, by investigating the AdS space, Hawking and Page \cite{Hawking} discovered that there is a phase transition between the thermal gas and stable large Schwarzschild black hole. Later, inspired by the AdS/CFT correspondence \cite{Maldacena,Gubser,Witten}, the well-known Hawking-Page phase transition was interpreted as the confinement/deconfinement phase transition in gauge theory \cite{Witten2}.

More interestingly, the phase transition can also occur between two different stable black hole phases in the same spacetime. For example, in the charged or rotating AdS black hole background, there was a phase transition between the stable small and large black holes \cite{Chamblin,Chamblin2,Caldarelli,Roychowdhury}. The study also implies that such phase transition is very similar to the liquid-gas phase transition of the van der Waals (VdW) fluid. On the other hand, for the charged AdS black holes, the similar phase transition and critical phenomena were also found in the $Q$-$\Phi$ (charge-electric potential) diagram \cite{Wu,Shen,Niu,Tsai,Wei,Zhou}. These results indicate that there exists a correspondence between the $Q$-$\Phi$ diagram of the charged AdS black hole and the $P$-$V$ (pressure-volume) diagram of the VdW fluid. However, there is a problem due to the fact that $Q$ and $\Phi$ are extensive and intensive quantities, while $P$ and $v$ are intensive and extensive quantities.

In order to solve the problem, finding the thermodynamic pressure $P$ is a key step. In fact, the solution of this question starts with the study of the cosmological constant $\Lambda$. Of particular interest, the cosmological constant was, recently, considered as a thermodynamic variable in the first law of black hole thermodynamics \cite{Kastor,Dolan,Dolan2,Dolan3,Cvetic,Pope,Kubiznak} for the reason that the cosmological constant is not fixed a priori in a theory but appears as vacuum expectation value and thus it can vary. Employing the consideration, the inconsistency between the first law of the thermodynamics and the Smarr relation was clarified \cite{Kastor} for a rotating AdS black hole. Meanwhile, we need to identify the black hole mass as the enthalpy rather than the internal energy of the system. Further, the cosmological constant was interpreted as the thermodynamic pressure \cite{Kastor,Gunasekaran}
\begin{eqnarray}
 P=-\frac{\Lambda}{8\pi}=\frac{(d-1)(d-2)}{16\pi l^{2}},\label{pL}
\end{eqnarray}
where $d$ is  the number of spacetime dimensions and $l$ is the AdS radius. Adopting such an interpretation, the $P$-$v$ criticality was first examined in detail in Ref. \cite{Kubiznak}. The result shows that the $P$-$v$ criticality, phase transition, and critical exponents are extremely similar to that of the VdW fluid. Subsequently, this study was extended to other AdS black holes, see Refs. \cite{Vahidinia,ChenLiuLiu,ZhaoZhao,MoLiu,ZouWang,XuZhaoZhao,WeiLiu,CaoXu,ChengWei,Teo,LiLi} and references therein. Remarkably, besides the phase transition of VdW type, some rich critical phenomena and phase transitions, such as the reentrant phase transition, triple point, isolated critical point, and $\lambda$-line phase transition were observed \cite{Altamirano,AltamiranoKubiznak,Altamirano3,Wei2,Frassino,Cai,XuZhao,
Kostouki,Hennigar,Hennigar2,Tjoa2,Ruihong}. These investigations not only give us new insight into the black hole thermodynamics, but also enhance our understanding on the cosmological constant.

Among the different AdS black holes, the rotating ones are more attractive. In a $d=$4 dimensional spacetime, a rotating Kerr AdS black hole demonstrates a small-large black hole phase transition of VdW type \cite{Gunasekaran}. And when the dimension of the spacetime increases, some novel phase transitions of great interesting occur. For $d\geq6$, besides the small-large black hole phase transition, the singly spinning Kerr AdS black hole also yields a reentrant phase transitions \cite{Altamirano}. In this case, the black hole can have a phase transition from a large black hole phase to a small one, and then back to the large black hole phase with the increase of the temperature or pressure. One of the phase transitions is first-order while another is zero-order. Moreover, for a six-dimensional multi-spinning Kerr AdS black hole, a triple point emerges in the phase diagram, which is very similar to the solid/liquid/gas phase transition of water \cite{AltamiranoKubiznak}. Nevertheless, the analytic or exact study of the phase transition for the rotating Kerr AdS black hole is very difficult. The reason is that, different from the charged AdS black hole case, the entropy of the rotating black hole depends not only on the horizon radius, but also on the angular momentum. Thus, only numerical result is possible.
	
From the other side, the black hole phase transition can be understood from the first law of thermodynamics. The condition determining the critical point was reexamined. The conventional condition that $\partial_{v}P=\partial_{v,v}P=0$ was extended to other cases. One useful condition is $\partial_{S}T=\partial_{S,S}T=0$ with $S$ the black hole entropy \cite{Spallucci,WeiLiu2}. Employing this condition, we obtained the analytic critical point for the four-dimensional Kerr AdS black hole \cite{Wei1}. For the higher-dimensional cases, the expression of the state equation becomes more complicated, and thus no exact result has been obtained. However, in the small angular momentum limit, the authors of Ref. \cite{Altamirano3} expanded the state equation and obtained an approximate analytic result. After examining the property of the thermodynamic quantities, we proposed a new viewpoint that all the thermodynamic quantities can be divided into two classes, the universal parameters and characteristic parameters \cite{WeiLiu2}. Adopting this point of view, it is natural to interpret the critical point as the relation between the universal and characteristic parameters. Further, the singly spinning Kerr AdS black hole was found to be a single-characteristic-parameter thermodynamic system, which can be used to determine the specific form of the critical point. Then the undetermined coefficients were obtained by using the critical condition and thus the exact critical points were derived. Furthermore, the exact critical reentrant phase transition points were also obtained for the rotating black holes in $d\geq6$ dimensions \cite{Wei1}.

Recently, higher dimensional rotating Kerr-AdS or Myers-Perry-AdS (MP-AdS) black hole has attracted great interest. Comparing with four dimensional Kerr-AdS black hole, five-dimensional and higher dimensional cases can possess more than one spinning parameters, which will lead to some novel phenomena and interest results. The Hamilton-Jacobi equation, massive Klein-Gordon equation, massive Dirac equation, and geodesic equation of the rotating MP black hole can be completely separated for the five and higher dimensional cases, and the two sets of equal spinning parameters enlarge the rotational symmetry group \cite{Vasudevan,Wuu,Demirchian,Salcedo}. The stability of five-dimensional Myers-Perry black holes with equal angular momenta were considered and several qualitative arguments were obtained in Ref. \cite{Murata,Kodama}. For two equal angular momenta, the scalar field perturbations were studied in five-dimensional spacetime in Refs. \cite{Zhidenko,Galajinsky}. It was found that the confining box can help the superradiant modes to extract rotational energy from the black hole. The shadow cast by five-dimensional MP black hole was studied \cite{Papnoi}. The upper bound of the radiation energy in the head-on collision of two MP black holes was found to significantly depend on the alignments of rotating axes for a given initial condition \cite{Gwak}. Following the gedanken experiment designed by Wald, it was found that the five-dimensional MP black hole cannot be over-spun and the weak cosmic censorship conjecture holds \cite{An}. All these studies indicate that higher dimensional rotating black holes are of great interest and worthy of being further investigated.

In particular, its thermodynamics was examined in Ref. \cite{Altamirano3,Pourdarvish}. In Ref. \cite{Altamirano3}, it was found that there exists a small-large black hole phase transition. Comparing with the singly spinning black hole, the thermodynamic phase transition and the critical phenomenon are closely dependent of both two spinning parameters for the five-dimensional Kerr-AdS black hole. This also enlarges the thermodynamic parameter space. Adopting the small angular momentum limit, the approximate critical point was obtained for the equal spinning black hole. In this paper, we would like to reconsider the thermodynamics for the multi-spinning Kerr-AdS black hole. Although the method we proposed in Ref. \cite{WeiLiu2} fails for the multi-spinning black hole case, it indeed works for the equal-spinning Kerr-AdS black hole because it is a single-characteristic-parameter thermodynamic system. In this paper, we mainly focus on the five-dimensional equal spinning Kerr-AdS black holes. The exact value of the critical point is obtained, which greatly improves the result given in Ref. \cite{Altamirano3}. Based on this result, we discuss the reduced thermodynamic law. The coexistence curve and the phase structure are also given. At last, the critical exponents near the exact critical point are numerically calculated when the system follows along the coexistence curve. Furthermore, a brief discussion and numerical study are devoted to the critical point of the unequal spinning black hole.

This paper is organized as follows. In Sec. \ref{bhg}, we show the black hole solution and its thermodynamic quantities. In the reduced parameter space, the thermodynamic laws and phase transition are discussed in detail in Sec. \ref{pspsa}. In Sec. \ref{psps}, the phase structure of the black hole system is displayed, respectively, in the $\tilde{P}$-$\tilde{T}$ diagram, $\tilde{T}$-$\tilde{V}$ diagram, and $\tilde{P}$-$\tilde{V}$ diagram. Different black hole phases are clearly illustrated. When the system follows along the coexistence curve, we calculate different critical exponents near the critical point in Sec. \ref{ce}. For the unequal spinning black hole, its exact critical point is also studied by employing the dimensional analysis and symmetry analysis in Sec. \ref{unequal}. Finally, we summarize and discuss our results in Sec. \ref{Conclusion}.

\section{Black hole and thermodynamic quantities}
\label{bhg}

A five-dimensional Kerr-AdS black hole possesses two spin parameters $a_{1}$ and $a_{2}$. Here, would like to consider the equal spinning case, i.e., $a_{1}=a_{2}=a$. Accordingly, in the Boyer-Lindquist coordinates, its metric is given by \cite{MyersPerry}
\begin{eqnarray}
 ds^{2}=&-&\frac{1}{\Xi}\left(1+\frac{r^{2}}{l^{2}}\right)dt^{2}
   +\frac{2m}{U\Xi^{2}}\left(dt-a(\mu_{1}^{2}d\phi_{1}+\mu_{2}^{2}d\phi_{2})\right)^{2}
   +\frac{U}{\Xi}(\mu_{1}^{2}d\phi_{1}^{2}+\mu_{2}^{2}d\phi_{2}^{2})\nonumber\\
   &+&\frac{Udr^{2}}{F-2m}
   +\frac{U}{\Xi}(d\mu_{1}^{2}+d\mu_{2}^{2})
   -\frac{U^2}{(r^{2}+l^{2})\Xi}
   \left(\mu_{1}d\mu_{1}+\mu_{2}d\mu_{2}\right)^{2},
\end{eqnarray}
where the metric functions read
\begin{eqnarray}
 && \Xi=1-\frac{a^{2}}{l^{2}},\quad
 U=r^{2}+a^{2},\\
 &&F=r^{-2}\left(1+\frac{r^{2}}{l^{2}}\right)(r^{2}+a^{2})^{2}.
\end{eqnarray}
Note that the coordinates $\mu_{1}$ and $\mu_{2}$ satisfy $\mu_{1}^{2}+\mu_{2}^{2}=1$. The thermodynamic quantities were obtained in Ref. \cite{GibbonsPerry}. For the five-dimensional equal spinning Kerr-AdS black hole, the black hole mass $M$, angular momentum $J$, and angular velocity $\Omega$ on the horizon are
\begin{eqnarray}
 M=\frac{\pi m(4-\Xi)}{4\Xi^{3}},\quad
 J=\frac{am\pi}{2\Xi^{3}},\quad
 \Omega=\frac{a(l^{2}+r_{h}^{2})}{l^{2}(r_{h}^{2}+a^{2})}.\label{angular2}
\end{eqnarray}
Here $r_{h}$ denotes the radius of the black hole horizon, which can be obtained by solving $F-2m=0$. The temperature $T$, entropy $S$, thermodynamic volume $V$ and specific volume $v$ read \cite{Cvetic,GibbonsPerry,Dolan8}
\begin{eqnarray}
 T&=&\frac{1}{2\pi}\left(\frac{2r_{h}(l^{2}+r_{h}^{2})}{l^{2}(a^{2}+r_{h}^{2})}-\frac{1}{r_{h}}\right),\label{temperaturee}\\
 S&=&\frac{A}{4}=\frac{\pi^{2}(a^{2}+r_{h}^{2})^{2}}{2r_{h}\Xi^{2}},\label{entropyy}\\
 V&=&\frac{r_{h}A}{4}+\frac{4\pi aJ}{3},\label{vv}\\
 v&=&\left(\frac{512V}{81\pi^{2}}\right),
\end{eqnarray}
where $A$ is the area of the black hole horizon. One can check this black hole solution obeys the following differential form
\begin{eqnarray}
 dM=TdS+2\Omega dJ+VdP,\label{fistlaw}
\end{eqnarray}
where the factor `2' comes from the two equal angular momenta $J_{1}=J_{2}=J$. of the black hole. In order to meet the first law of the ordinary thermodynamic fluid, the black hole mass $M$ needs to be treated as the enthalpy $H$ rather the energy of the black hole. Then the Gibbs free energy will be
\begin{eqnarray}
 G=H-TS=\frac{3\pi(a^{2}+r_{h}^{2})}{8r_{h}^{2}(3-4\pi a^{2}P)^{3}}
 \bigg(9r_{h}^{2}-12\pi P r_{h}^{4}+4\pi P a^{4}(4\pi P r_{h}^{2}-3)+a^{2}(45+8\pi P r_{h}^{2}(9+10\pi Pr_{h}^{2}))\bigg).
\end{eqnarray}

\section{Reduced thermodynamics}
\label{pspsa}

In this section, we would like to study the thermodynamic laws and phase transition for the black hole in the reduced parameter space. As shown in Ref. \cite{Altamirano3}, there exists a VdW like phase transition. According to the dimensional analysis \cite{Wei1}, the critical point must have the following relations
\begin{eqnarray}
 P_{c}\sim J^{-\frac{2}{3}},\quad
 T_{c}\sim J^{-\frac{1}{3}},\quad
 v_{c}\sim J^{\frac{1}{3}},\label{form}
\end{eqnarray}
In the small angular momentum limit, the critical values have been obtained in Eq. (102) of Ref. \cite{Altamirano3} for $d$-dimensional equal spinning Kerr-AdS black holes. Taking $d$=5, one can obtain the following result:
\begin{eqnarray}
 P_{c}^{a}=0.0181J^{-\frac{2}{3}},\quad
 T_{c}^{a}=0.1227J^{-\frac{1}{3}},\quad
 v_{c}^{a}=2.9655J^{\frac{1}{3}}.\label{formmm}
\end{eqnarray}
It is obvious that it has the same form as (\ref{form}) given by the dimensional analysis. Then we can reduce the thermodynamic quantities with their critical values. For example, the reduced pressure and temperature are defined as
\begin{eqnarray}
 \tilde{P}=\frac{P}{P_{c}}, \quad
 \tilde{T}=\frac{T}{T_{c}}.
\end{eqnarray}
After this, the critical point will be shifted to (1, 1). In the following, it is interesting to discuss the black hole thermodynamics in the reduced parameter space. For simplicity, we fix the angular momentum $J$. Then the first law (\ref{fistlaw}) will be of the form
\begin{eqnarray}
 d\tilde{H}&=&R_{c1}\tilde{T}d\tilde{S}+R_{c2}\tilde{V}d\tilde{P},\\
 d\tilde{G}&=&-R_{c3}\tilde{S}d\tilde{T}+R_{c4}\tilde{V}d\tilde{P},\label{fistlaw2}
\end{eqnarray}
where the critical ratios $R_{ci}$ ($i=1,2,3,4$) are dimensionless constants, and they are
\begin{eqnarray}
 R_{c1}=\frac{T_{c}S_{c}}{H_{c}},\quad
 R_{c2}=\frac{P_{c}V_{c}}{H_{c}},\quad
 R_{c3}=\frac{T_{c}S_{c}}{G_{c}},\quad
 R_{c4}=\frac{P_{c}V_{c}}{G_{c}}.
\end{eqnarray}

\subsection{Maxwell equal area laws}

As we know, the black hole phase transition point can be exactly determined by the swallow tail behavior of the Gibbs free energy. Alternatively, one can also construct two equal areas on each isothermal line or isobaric line in $P$-$V$ or $T$-$S$ plane by drawing a horizontal line. It is worthwhile pointing out that the equal area law does not hold in some cases, for details, see Ref. \cite{WeiLiu2}.

In ordinary parameter space, we can get the Maxwell equal area laws from the first law for fixed temperature and pressure, respectively,
\begin{eqnarray}
 \int_{P_{\ast}}^{P_{\ast}} VdP&=&0,\\
 \int_{T_{\ast}}^{T_{\ast}} SdT&=&0,
\end{eqnarray}
where $P_{\ast}$ and $T_{\ast}$ denote the pressure and temperature of the phase transition. Note that the above two integrals should be calculated along the isothermal line and isobaric line, respectively. The lower and upper integration limits denote different states of the black hole system, and thus these integrals are nonzero. Since $R_{ci}$ are dimensionless constants, we can find that the equal area laws are also held in the reduced parameter space. For fixed reduced temperature and pressure, they are
\begin{eqnarray}
 \int_{\tilde{P}_{\ast}}^{\tilde{P}_{\ast}} \tilde{V}d\tilde{P}&=&0,\\
 \int_{\tilde{T}_{\ast}}^{\tilde{T}_{\ast}} \tilde{S}d\tilde{T}&=&0.
\end{eqnarray}
So in the reduced parameter space, the Maxwell equal area laws have the similar forms, and one can use them to determine the reduced phase transition point. Moreover, at the phase transition point of the small and large black hole phases, we have
\begin{eqnarray}
 \Delta\tilde{G}=\tilde{G}_{l}(\tilde{T}_{\ast}, \tilde{P}_{\ast})-\tilde{G}_{s}(\tilde{T}_{\ast}, \tilde{P}_{\ast})=0.
\end{eqnarray}
Here $\tilde{G}_{l}$ and $\tilde{G}_{s}$ are the reduced Gibbs free energy for the small and large black holes, respectively. Hence, it is natural that the swallow tail behavior also exists for the reduced Gibbs free energy in the reduced parameter space.

\subsection{Clapeyron equation}

Clapeyron equation gives the slope of the coexistence curve in the $P$-$T$ diagram. Its generalizations for the charged AdS black hole and rotating AdS black hole have been done in Refs. \cite{WeiLiu2,Wei1}. Here we only consider the conventional one, which is
\begin{eqnarray}
 \left(\frac{dP}{dT}\right)_{J}=\frac{\Delta S}{\Delta V}.
\end{eqnarray}
By using Eq. (\ref{fistlaw2}), we can obtain the reduced Clapeyron equation:
\begin{eqnarray}
 \left(\frac{d\tilde{P}}{d\tilde{T}}\right)_{J}=\frac{R_{c3}}{R_{c4}}\cdot\frac{\Delta \tilde{S}}{\Delta \tilde{V}}.\label{Clapeyronq}
\end{eqnarray}
During the small-large black hole phase transition, there will be latent heat absorbed or released by the system. In the reduced parameter space, the reduced latent heat can be calculated with
\begin{eqnarray}
 \tilde{L}=\frac{L}{T_{c}S_{c}}=\tilde{T}\Delta\tilde{S}.
\end{eqnarray}

\subsection{Heat capacity}

Heat capacity often marks the local instability of a thermodynamic system. Positive and negative heat capacities correspond to stable and unstable systems, respectively. In the reduced parameter space, we can define the reduced heat capacity as
\begin{eqnarray}
 \tilde{C}_{\tilde{J}, \tilde{P}}=\frac{C_{J, P}}{S_{c}}
   =\tilde{T}\left(\frac{d\tilde{S}}{d\tilde{T}}\right)_{\tilde{J}, \tilde{P}}.\label{heatcap}
\end{eqnarray}
Since the critical value $S_{c}$ is positive, $\tilde{C}_{\tilde{J}, \tilde{P}}$ and $C_{J, P}$ have the same sign. Then, we can use the reduced heat to explore the instability for the black hole system in the reduced parameter space.

\section{Phase transition and phase structures}
\label{psps}

In this section, we would like to consider the black hole phase transition and phase structure for the five-dimensional equal spinning Kerr-AdS black hole.

\begin{figure}
\center{\subfigure[]{\label{pTrps610}
\includegraphics[width=6cm]{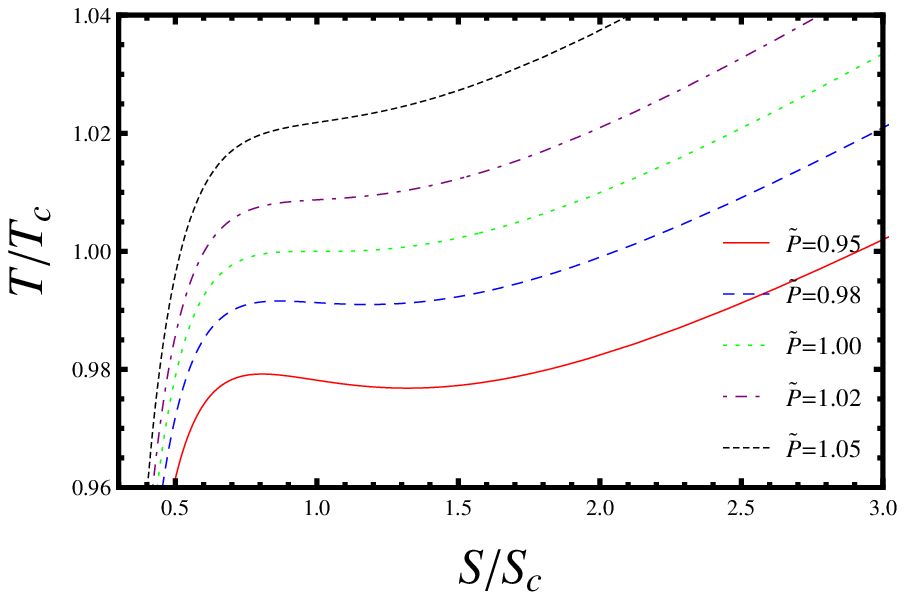}}
\subfigure[]{\label{pTups610}
\includegraphics[width=6cm]{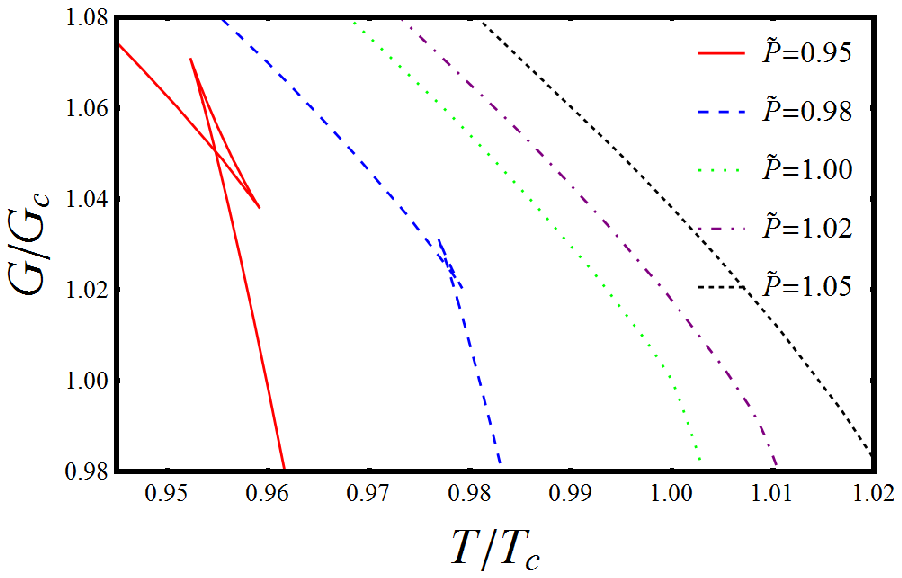}}}
\caption{Isobaric lines in the reduced parameter space for fixed reduced pressure $\tilde{P}$=0.95, 0.98, 1, 1.02, and 1.05. (a) $\tilde{T}$-$\tilde{S}$ diagram. The fixed pressure increases from bottom to top. (b) $\tilde{G}$-$\tilde{T}$ diagram. The fixed pressure increases from left to right.}\label{ptsgt}
\end{figure}

At first step, we focus on the exact critical point. As discussed in Ref. \cite{WeiLiu2}, it is easy to determine the point by using
\begin{eqnarray}
 (\partial_{S}T)_{J, P}=(\partial_{S,S}T)_{J, P}=0,\label{ccc}
\end{eqnarray}
for the rotating AdS black hole. Since the temperature given in (\ref{temperaturee}) is a function of $l$, $a$, and $r_{h}$, we need to change these parameters to $P$, $J$, and $S$. Here we summarize how to do this. Using Eq. (\ref{pL}), we can change the parameter $l$ to $P$. Then by solving the entropy equation showed in (\ref{entropyy}), we have
\begin{eqnarray}
 a^{2}=\frac{3 \left(8 P S r_h\pm\sqrt{2} \left(4 \pi  P
   r_h^2+3\right) \sqrt{S r_h}+3 \pi
   r_h^2\right)}{\pi  \left(32 P^2 S r_h-9\right)}.
\end{eqnarray}
Adopting the `+' sign in the equation, we will get negative temperature $T=-16P^{2}S/9\pi$, which corresponds to the naked singularity, and thus we will not consider this solution. Then choosing the negative sign and plunging it into the angular momentum formula (\ref{angular2}), one can express the horizon radius as $r_{h}=r_{h}(J, P, S)$. Finally, substituting all these results into the temperature (\ref{temperaturee}), we can obtain $T=T(J, P, S)$. Adopting the condition (\ref{ccc}), we get the exact critical point
\begin{eqnarray}
 P_{c}=0.0229J^{-\frac{2}{3}},\quad
 T_{c}=0.1381J^{-\frac{1}{3}},\quad
 S_{c}=37.8149J,\quad
 V_{c}=74.5554J^{\frac{4}{3}},\quad
 v_{c}=2.6287J^{\frac{1}{3}}.
\end{eqnarray}
It is worthwhile noting that this result has no approximation of the angular momentum. The result has a small difference between these given in (\ref{formmm}) obtained in Ref. \cite{Altamirano3} by taking small angular momentum limit. Comparing our result with them, we obtain the relative deviations
\begin{eqnarray}
\Delta P_{c}\approx21\%,\quad \Delta T_{c}\approx11\%, \quad \Delta v_{c}\approx-13\%,\quad \Delta \left(\frac{P_{c}v_{c}}{T_{c}}\right)\approx0.4\%.
\end{eqnarray}
It is clear that there is a large deviation between the our exact result and the approximate one. Another significant feature is that these relative deviations are independent of the angular momentum $J$, which is also similar to the $d$-dimensional singly spinning Kerr-AdS black hole cases \cite{Wei1}. The reason of this is mainly because that the five-dimensional equal spinning Kerr-AdS black hole is a single-characteristic-parameter systems. Moreover, in Ref. \cite{Altamirano3}, they kept the small quantity $(\frac{a}{l})$ to the fourth order in the state equation. We believe that if higher orders are included in, the approximate result (\ref{formmm}) will approach our exact one. For the unequal spinning black hole case, $J_{1}\neq J_{2}$, we discuss and give the exact values of the critical points in Sec. \ref{unequal}. The result also shows that the critical point is only dependent on $\epsilon=J_{2}/J_{1}\in(0, 1)$. Moreover, we calculate the four dimensionless critical ratios
\begin{eqnarray}
 R_{c1}=0.8192,\quad
 R_{c2}=0.2683,\quad
 R_{c3}=4.5320,\quad
 R_{c4}=1.4840.\quad
\end{eqnarray}
After obtaining the values of the critical point, we show the reduced isobaric lines in Fig. \ref{ptsgt} for fixed $\tilde{P}$=0.95, 0.98, 1, 1.02, and 1.05. In Fig. \ref{pTrps610}, the isobaric lines are displayed in the $\tilde{T}$-$\tilde{S}$ diagram. From it, we can find that for $\tilde{P}<$1, there exists a non-monotonic behavior for each curve, and two extremal points are presented. As $\tilde{P}$ increases, these two extremal points approach to each other, and coincide at the critical point. When $\tilde{P}>$1, the non-monotonic behavior disappears. Thus, the reduced temperature monotonously increases with the reduced entropy. In fact, the non-monotonic behavior of the isobaric line implies the existence of the phase transition. We can construct the equal area law to determine the phase transition point. In the $\tilde{G}$-$\tilde{T}$ diagram, for the reduced isobaric lines with $\tilde{P}<1$, the swallow tail behaviors are presented. Considering that the Gibbs free energy prefers a phase of low $\tilde{G}$, the phase transition takes place at the intersection point of the two different branches of the system phases.

\begin{figure}
\centerline{\includegraphics[width=6cm]{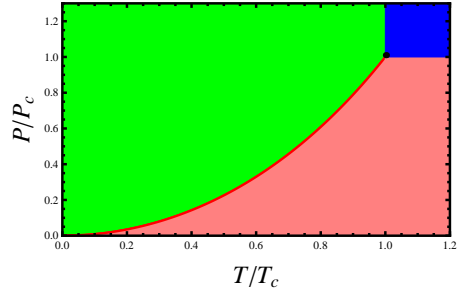}}
\caption{Phase structure in  $\tilde{P}$-$\tilde{T}$ diagram.} \label{Pptstructure}
\end{figure}

Here, we adopt the latter one to determine the phase transition point. Through varying the pressure, one can obtain the temperature of the phase transition accordingly. In Fig. \ref{Pptstructure}, we describe the phase structure in the $\tilde{P}$-$\tilde{T}$ diagram. The red thick line is the coexistence curve of small and large black holes. The black dot denotes the critical point. The green region above the coexistence curve is for the small black hole phase. And the red region below the curve is for the large black hole phase. The blue region located at the upper right corner is for the supercritical phase, within which the small and large black hole will not be clearly distinguished.

Coexistence curve is an important key to study the physical change among the phase transition. However, for this case, there is no analytic form. Fortunately, we can adopt the fitting method used in Ref. \cite{WeiLiu2} to obtain a highly
accurate fitting formula for the coexistence curve. The result is
\begin{eqnarray}
 \tilde{P}&=&0.870469\tilde{T}^2-0.000219 \tilde{T}^3+0.136996 \tilde{T}^4
      -0.057063\tilde{T}^5+0.079964 \tilde{T}^6\nonumber\\
      &&-0.057912 \tilde{T}^7+0.039096\tilde{T}^8
      -0.013912 \tilde{T}^9+0.002581 \tilde{T}^{10}.\label{pot}
\end{eqnarray}
For low $\tilde{T}$, the relative deviation is less than 0.002\%. And for high $\tilde{T}$, the relative deviation can reach $10^{-8}$.

\begin{figure}
\center{\subfigure[]{\label{pEpts}
\includegraphics[width=6cm]{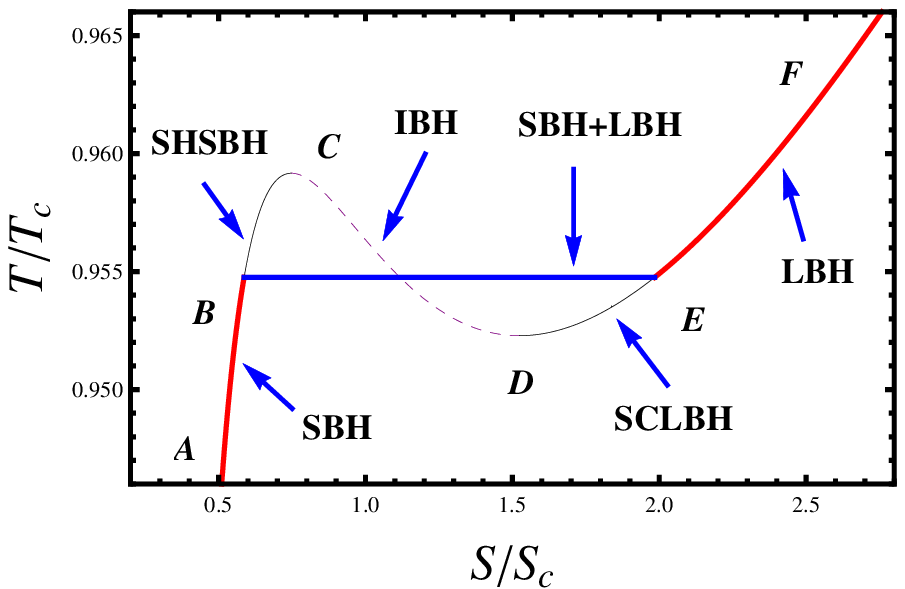}}
\subfigure[]{\label{pEpgt}
\includegraphics[width=6cm]{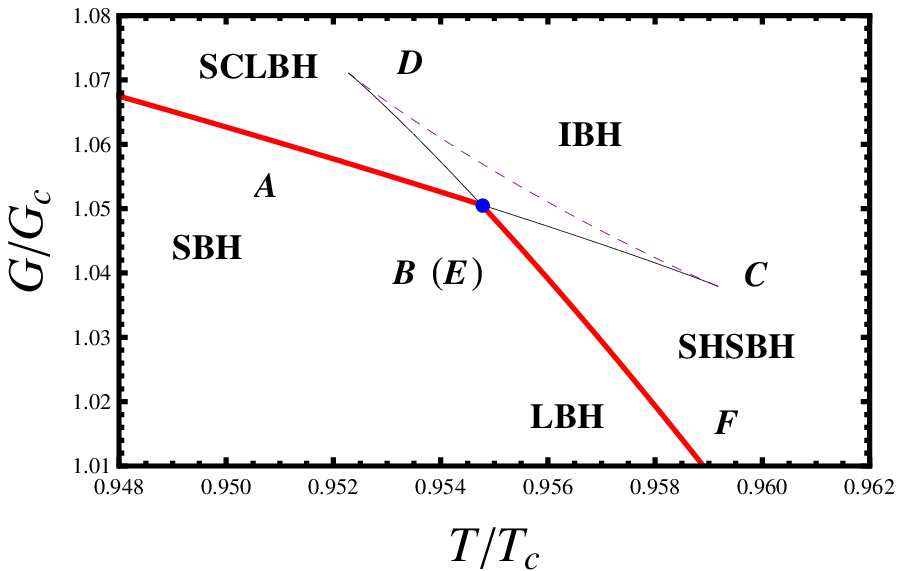}}}
\caption{Isobaric line in the reduced parameter space for fixed reduced pressure $\tilde{P}$=0.9. The solid lines have positive heat capacity, while the dashed ones have negative capacity. (a) $\tilde{T}$-$\tilde{S}$ diagram. (b) $\tilde{G}$-$\tilde{T}$ diagram.}\label{ppEpgt}
\end{figure}

Next, we would like to turn to construct the phase structure in the $\tilde{P}$-$\tilde{V}$ and $\tilde{T}$-$\tilde{V}$ diagrams. Before doing it, we would like to give a brief note on the metastable phase. For example, we plot the isobaric line with $\tilde{P}=0.9$ in Fig. \ref{ppEpgt} by using the state equation. In Fig. \ref{pEpts}, the isobaric line is divided into five branches: AB, BC, CD, DE, and EF, which are named as the small black hole (SBH) branch, superheated small black hole (SHSBH) branch, intermediate black hole (IBH) branch, supercooled large black hole (SCLBH) branch, and large black hole (LBH) branch, respectively. Among these branches, only the intermediate black hole branch has negative heat capacity, which indicates thermodynamic instability. According to the thermodynamics, the phase transition occurs at $\tilde{T}$=0.9548. So the curve BCDE will be replaced by the horizontal line BE. Then with the increase of the entropy or the black hole horizon radius, the system will encounter the small black hole branch AB, coexistence black hole branch BE, and large black hole branch EF. This can also be found in Fig. \ref{pEpgt}. In addition, due to the density fluctuations or interactions, the system can reach the superheated small black hole branch BC with the increase of the black hole entropy, or reach the supercooled large black hole branch DE with the decrease of the black hole entropy. Nevertheless, these two branches are two metastable phases of the system. In the phase diagram Fig. \ref{Pptstructure}, the two metastable phases are not shown. However, in another parameter space, we can clearly display them.

\begin{figure}
\center{\subfigure[]{\label{ptvstructure}
\includegraphics[width=6cm]{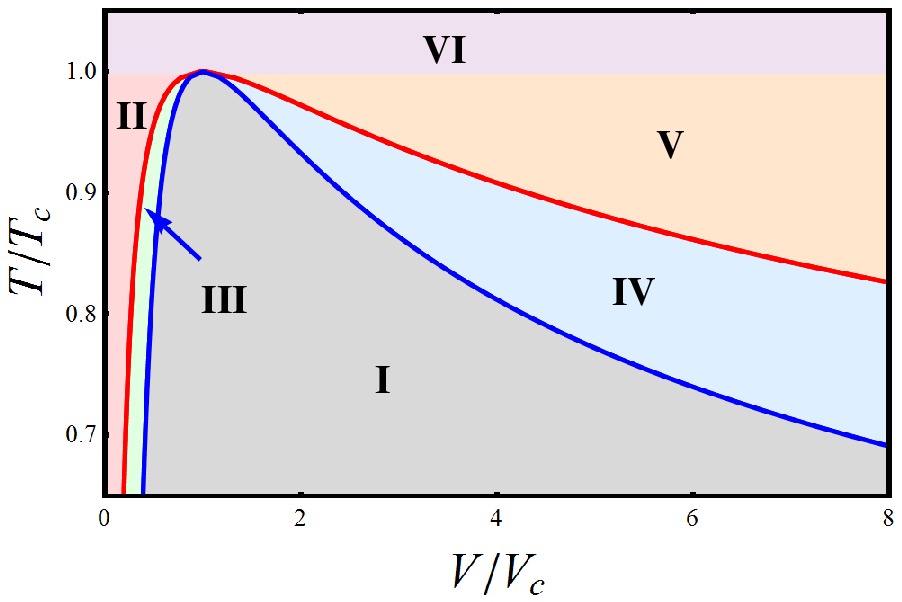}}
\subfigure[]{\label{pEpgtd}
\includegraphics[width=6cm]{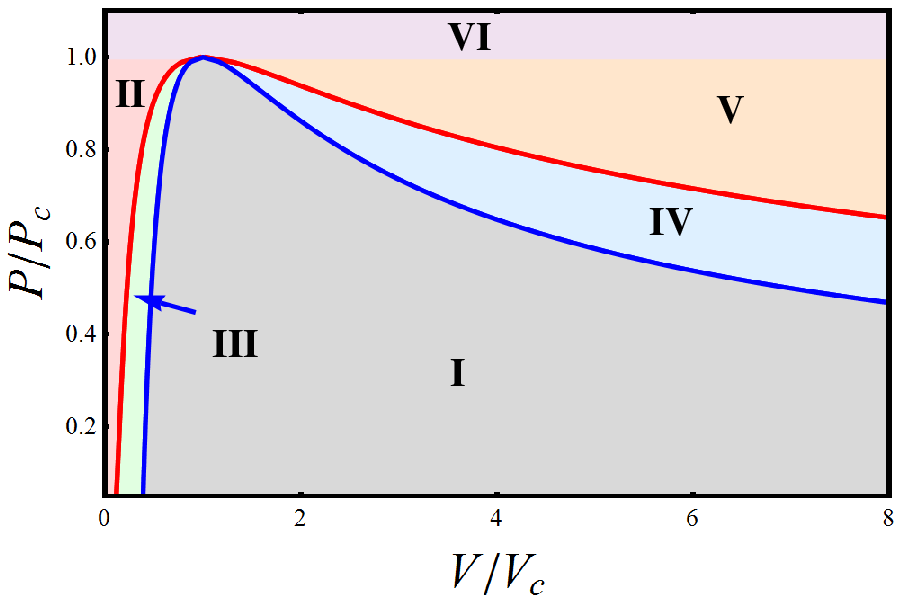}}}
\caption{Phase structures for the black hole system. (a) $\tilde{T}$-$\tilde{V}$ diagram. (b) $\tilde{P}$-$\tilde{V}$ diagram.}\label{ppEpgdt}
\end{figure}

We show the phase structures of the black hole in the $\tilde{T}$-$\tilde{V}$ diagram and $\tilde{P}$-$\tilde{V}$ diagram in Fig. \ref{ppEpgdt}. The phase structures are similar. Six phases are presented. Regions I$\sim$VI are, respectively, the small-large black hole coexistence phase, small black hole phase, superheated small black hole phase, supercooled large black hole phase, large black hole phase, and supercritical black hole phase. Among these phases, Regions III and IV are two metastable phases. It is also worthwhile noting that in the coexistence region I, the state equation does not hold.

\begin{figure}
\centerline{\includegraphics[width=6cm]{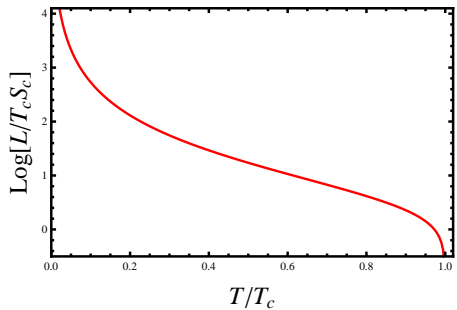}}
\caption{Behavior of the reduced latent heat.} \label{PLatentheat}
\end{figure}

On the other hand, employing the fitting formula (\ref{pot}), we can check the reduced Clapeyron equation given in (\ref{Clapeyronq}). Moreover, the reduced latent heat can also be calculated and the result is shown in Fig. \ref{PLatentheat}. It is clearly that the reduced latent heat decreases with the temperature of the phase transition. At low temperature, the reduced latent heat has a very large value, which is of order $10^4$. However, when the critical temperature is approached, the latent heat vanishes, indicating a second order phase transition. Note that at the critical point, $\tilde{L}=0$, and thus one has $\log\tilde{L}=-\infty$, however we have not clearly show it in Fig. \ref{PLatentheat}.

\section{Critical exponents}
\label{ce}

Near the critical point, there is a number of interesting behaviors. For example, the heat capacity defined in (\ref{heatcap}) diverges at the critical point. And the microstructures of the small and large black holes also approach the same \cite{WeiLiu}. In particular, the critical exponents can reflect certain universality of the phase transition. So, here we would like to examine the critical exponents near the critical point.

We first examine the critical exponent for the heat capacity. When the black hole system approaches the critical point along the coexistence curve, we can obtain the following result with the fitting method of the numerical result
\begin{equation}
 \tilde{C}_{\tilde{J}, \tilde{P}}=\begin{cases}
 0.9798\times|1-\tilde{T}|^{-1.0411},&  $for coexistence small black hole$,\\
 2.3189\times|1-\tilde{T}|^{-0.9576},&  $for coexistence large black hole$.\\
\end{cases}
\end{equation}
So, the critical exponent approximately equals to 1 for both the coexistence small and large black holes. This result is also the same as that of Ref. \cite{LiangWei}. In addition, we have another exponent
\begin{equation}
 (1- \tilde{P})=\begin{cases}
 0.1142\times|1-\tilde{V}|^{1.9058},&  $for coexistence small black hole$,\\
 0.2318\times|1-\tilde{V}|^{2.0954},&  $for coexistence large black hole$.\\
\end{cases}
\end{equation}
Considering the error of calculation and fitting formula, we expect that the exponent is 2.

On the other hand, the microstructures of the small and large black holes approach the same. Thus it is interesting to consider other exponents. Apply the similar process, we have
\begin{eqnarray}
 \Delta\tilde{S}&=&(\tilde{S}_{L}-\tilde{S}_{S})
  \sim 5.7123\times|1-\tilde{T}|^{0.5014},\\
 \Delta\tilde{V}&=&(\tilde{V}_{L}-\tilde{V}_{S})
  \sim 7.6606\times|1-\tilde{T}|^{0.5020},\\
 \Delta\tilde{v}&=&(\tilde{v}_{L}-\tilde{v}_{S})
  \sim 1.8876\times|1-\tilde{T}|^{0.5004}.
\end{eqnarray}
It is clearly that near the critical point, $\Delta\tilde{S}$, $\Delta\tilde{V}$, and $\Delta\tilde{v}$ exactly have the same critical exponent $\frac{1}{2}$. As we know, the order parameter has a critical exponent $\frac{1}{2}$. So we can treat them as the order parameter to describe the small-large black hole phase transition. From the above result, one gets
\begin{eqnarray}
 \frac{\Delta\tilde{S}}{\Delta\tilde{V}}=0.7457,
\end{eqnarray}
which means that near the critical point, $\Delta\tilde{S}$ and $\Delta\tilde{V}$ have a linear relation. Further, the slope (\ref{Clapeyronq}) of the coexistence curve near the critical point is
\begin{eqnarray}
 \left(\frac{d\tilde{P}}{d\tilde{T}}\right)_{J}=2.2772.
\end{eqnarray}
One needs to note that the factors $R_{c3}$ and $R_{c4}$ have been included in. In fact, we can also obtain the slope from the fitting formula (\ref{pot}), $\left(\frac{d\tilde{P}}{d\tilde{T}}\right)_{J}=2.2907$ near the critical point. The relative deviation between these two results is about 0.6\%. On the other hand, at other temperature of the phase transition, the slope can be obtained directly with (\ref{pot}).

\section{Critical point for unequal spinning black holes}
\label{unequal}

As shown above, there are two angular momenta $J_{1}$ and $J_{2}$ in the spacetime. If both $J_{1}$ and $J_{2}$ are nonzero and unequal, the critical phenomena will be quite complicated. However, some interesting properties can be obtained by simple dimensional analysis and symmetry analysis.

Both the angular momenta $J_{1}$ and $J_{2}$ can take values form $(0, \infty)$, where we only require the angular momentum is positive. On the other hand, $J_{1}$ and $J_{2}$ are symmetrical, so the thermodynamics is invariable under
\begin{eqnarray}
 J_{1}\leftrightarrow J_{2}.\label{symmetrical}
\end{eqnarray}
Considering this symmetry, we would like to reduce the thermodynamic state parameter with $X=\sqrt{J_{1}J_{2}}$. For example, a reduced state parameter $A$, such as the temperature $T$ and pressure $P$, reads
\begin{eqnarray}
 \tilde{A}=\frac{A}{X^{\alpha}},
\end{eqnarray}
with $\alpha$ being the dimension number of $A$ measured with the angular momentum $J$. The specific values can be found in Ref. \cite{Wei1}. For simplicity, we define a dimensionless ratio of the angular momenta
\begin{eqnarray}
 \epsilon=\frac{J_{2}}{J_{1}}, \quad \epsilon\in(0, \infty).
\end{eqnarray}
For the unequal spinning MP-AdS black hole, the state equation has an abstract form
\begin{eqnarray}
 T=T(P,\;V,\;J_{1},\;J_{2}).
\end{eqnarray}
After reducing this state equation with parameter $X$, one will get
\begin{eqnarray}
 \tilde{T}=\tilde{T}(\tilde{P},\;\tilde{V},\;\frac{1}{\sqrt{\epsilon}},\; \sqrt{\epsilon}).
\end{eqnarray}
Further, we can reexpress the reduced temperature as
\begin{eqnarray}
 \tilde{T}=\tilde{T}(\tilde{P},\;\tilde{V},\;\epsilon),
\end{eqnarray}
which means that the reduced temperature is a function of the reduced pressure $\tilde{P}$, volume $\tilde{V}$, and the dimensionless parameter $\epsilon$.

On the other hand, since the existence of the symmetry equation (\ref{symmetrical}), the reduced temperature $\tilde{T}$ depends on the same form of $\frac{1}{\sqrt{\epsilon}}$ and $\sqrt{\epsilon}$, which leads to an interesting result that the reduced thermodynamics is invariable under
\begin{eqnarray}
 \epsilon\leftrightarrow\frac{1}{\epsilon}.
\end{eqnarray}
Therefore, we only need to consider $\epsilon\in(0,\;1)$. Based on the above analysis, we can obtain the result that the critical phenomena and phase transition for the unequal spinning Kerr-AdS black holes are only dependent of the dimensionless parameter $\epsilon\in(0,\;1)$ in the reduced parameter space.

Here we show the specific values of the reduced critical point in Table \ref{tab1}. For a given critical point, we can study the coexistence curve, phase diagram, and critical exponents for the unequal spinning Kerr-AdS black hole. However, the results are similar to these of the equal spinning black hole.

\begin{table}[h]
\begin{center}
\begin{tabular}{cccccc}
  \hline\hline
 $\epsilon$& $\tilde{P}_{c}$ & $\tilde{T}_{c}$ & $\tilde{S}_{c}$ & $\tilde{V}_{c}$ & $\tilde{r}_{hc}$ \\\hline
 0.1 & 0.013599 & 0.106259 & 80.9250 & 205.641 & 2.5041 \\
 0.2 & 0.016930 & 0.118572 & 58.4345 & 133.214 & 2.2469 \\
 0.3 & 0.019023 & 0.125698 & 49.2674 & 106.102 & 2.1231 \\
 0.4 & 0.020442 & 0.130316 & 44.4168 & 92.4045 & 2.0515 \\
 0.5 & 0.021416 & 0.133396 & 41.5796 & 84.6173 & 2.0072 \\
 0.6 & 0.022074 & 0.135440 & 39.8542 & 79.9666 & 1.9794 \\
 0.7 & 0.022501 & 0.136753 & 38.8073 & 77.1768 & 1.9621 \\
 0.8 & 0.022759 & 0.137539 & 38.2027 & 75.5773 & 1.9520 \\
 0.9 & 0.022891 & 0.137939 & 37.9013 & 74.7829 & 1.9469 \\
 1.0 & 0.022929 & 0.138055 & 37.8149 & 74.5554 & 1.9455 \\
\hline\hline
\end{tabular}
\caption{Specific values of the reduced critical point for the unequal spinning MP-AdS black hole with $\epsilon$=0.1$\sim$ 1.0. The values are also invariable under $\epsilon\leftrightarrow\frac{1}{\epsilon}$.}\label{tab1}
\end{center}
\end{table}

\section{Conclusions}
\label{Conclusion}

In this paper, we studied the critical phenomena and phase structures for the five-dimensional equal spinning Kerr-AdS black hole. This system demonstrates a small-large black hole phase transition of VdW type.

For the equal spinning black hole, by employing with the dimensional analysis, we obtained the exact values of the critical points of the phase transition and greatly improved the result given in Ref. \cite{Altamirano3}, where the small angular momentum limit was used.

We examined the reduced thermodynamic laws and introduced several dimensionless critical rations. Based on them, we considered the condition to determine the phase transition point. The result states that, in the reduced parameter space, the generalized Maxwell equal area laws $\int\tilde{V}d\tilde{P}=0$ and $\int\tilde{S}d\tilde{T}=0$ hold. And the phase transition point can also be determined by the swallow tail behavior of the Gibbs free energy. Moreover, the Clapeyron equation was also modified in the reduced parameter space. The reduced latent heat and heat capacity were also calculated.

Then based on the reduced thermodynamic laws, we investigated the phase structures. In the $\tilde{P}$-$\tilde{T}$ diagram, $\tilde{T}$-$\tilde{V}$ diagram, and $\tilde{P}$-$\tilde{V}$ diagram, the phase structures were clearly plotted. Especially, in the $\tilde{T}$-$\tilde{V}$ and $\tilde{P}$-$\tilde{V}$ diagrams, two new metastable phases, a superheated small black hole phase and a supercooled large black hole phase, were shown. The coexistence curve in the $\tilde{P}$-$\tilde{T}$ diagram is a monotonically increasing function of temperature, and ends with the reduced critical point located at (1, 1). By combining with the numerical result, we obtained a highly accurate fitting formula for the coexistence curve, which is very useful on studying the black hole phase transition. For example, by employing this formula, one can check the reduced Clapeyron equation. The latent heat can also be calculated, see Fig. \ref{PLatentheat}.

Considering that the black hole system varies along the coexistence small and large black hole curve, we numerically calculated the critical exponents. It was found that the heat capacity has an exponent 1. Moreover, all $\Delta\tilde{S}$, $\Delta\tilde{V}$, and $\Delta\tilde{v}$ have exponent $\frac{1}{2}$, which implies that all these quantities can act as order parameter to describe the small-large black hole phase transition.

Additionally, we investigated the critical point for the unequal spinning Kerr-AdS black hole. Through reducing all the thermodynamic quantities with $\sqrt{J_{1}J_{2}}$ and introducing a dimensionless ratio $\epsilon=J_{2}/J_{1}$, we found that the reduced critical point only depends on $\epsilon$, and is invariable under $\epsilon\leftrightarrow\frac{1}{\epsilon}$. Then we listed some exact values of the critical point for $\epsilon\in(0, 1)$. These results are critical for studying the property of the MP-AdS black hole system among the phase transition.

\section*{Acknowledgements}
We would like to thank Robert B. Mann for useful discussions. This work was supported by the National Natural Science Foundation of China (Grants No. 11675064, No. 11875151, and No. 11522541). S.-W. Wei was also supported by the Chinese Scholarship Council (CSC) Scholarship (201806185016) to visit the University of Waterloo.

\end{document}